\documentclass[3p,times]{elsarticle}

\usepackage{ecrc}

\volume{00}

\firstpage{1}

\journalname{Nuclear Physics A}

\runauth{J. Vijande {\it et al.}}

\jid{NUPHA}

\jnltitlelogo{Nuclear Physics A}

\CopyrightLine{2012}{Published by Elsevier Ltd.}

\usepackage{amssymb}
\usepackage{amsthm}
\def\ka{\left.\mid 11 \right\rangle_c}
\def\kb{\left.\mid 88 \right\rangle_c}
\def\kap{\left.\mid 1'1' \right\rangle_c}
\def\kbp{\left.\mid 8'8' \right\rangle_c}
 \usepackage{graphics}
\usepackage{epsfig}
\usepackage{grffile}

\begin{document}

\begin{frontmatter}

\dochead{}



\title{Heavy hadron spectroscopy: a quark model perspective}

\author[1]{J. Vijande}
\address[1]{Departamento de F\'{\i}sica At\'{o}mica, Molecular y Nuclear, Universidad de Valencia (UV)
and IFIC (UV-CSIC), Valencia, Spain.}

\author[2]{A. Valcarce}
\author[2]{T.F. Caram\'es}
\address[2]{Departamento de F{\'\i}sica Fundamental,
Universidad de Salamanca, 37008 Salamanca, Spain}

\author[3]{H. Garcilazo}
\address[3]{Escuela Superior de F\'\i sica y Matem\'aticas,
Instituto Polit\'ecnico Nacional,
Edificio 9, 07738 M\'exico D.F., Mexico}

\begin{abstract}
We present recent results of hadron spectroscopy and hadron-hadron interaction from the perspective of constituent quark models.
We pay special attention to the role played by higher order Fock space components in the hadron spectra and the
connection of this extension with the hadron-hadron interaction. The main goal of our description is
to obtain a coherent understanding of the low-energy hadron phenomenology without enforcing any particular
model, to constrain its characteristics and learn about low-energy realization of the theory.
\end{abstract}

\begin{keyword}
Heavy hadrons \sep Constituent quark model \sep Many-quark systems
\end{keyword}

\end{frontmatter}



\section{Introduction}
\label{introduction}

For almost thirty years after the discovery of the $J/\psi$ and its excitations, the so-called 
November revolution~\cite{Bjo85}, heavy hadron spectroscopy 
was at rest. Paraphrasing Lord Kelvin famous speech~\cite{Kel01}, by 2003 there were only two clouds on the horizon obscuring 
{\it the beauty and clearness of the dynamical theory}. In the hadron spectra these {\it two clouds} were on the one hand
the missing resonance problem, i.e., all quark models predict
a proliferation of excited states which have not been measured, and, on the other hand, the observation by BaBar
of an open-charm meson, the $D_{sJ}^*(2317)$, whose properties were quite different 
from those predicted by quark potential models.

Since those peaceful days an increase on both amount and quality of 
experimental data has shown quite a different picture, far more involved and convoluted, with 
the observation of the well-known $X(3872)$, the $X(3940)$, $Y(3940)$, $Z(3930)$, $Y(4140)$, and several other states.
Based on Gell-Mann conjecture~\cite{Gel64} the hadronic experimental data were classified either as $q\overline{q}$ or $qqq$
states according to $SU(3)$ irreducible representations, nowadays this hypothesis may be in question. 
Therefore, the study of the role played by higher order Fock space components in the hadron spectra, allowed by the Gell-Mann
classification, is an interesting issue to address. 

In this talk we give an overview of a project for getting a coherent
understanding of the low-energy hadron phenomenology from the perspective of constituent quark models~\cite{Val05}. 
We will make emphasis on three different aspects. Firstly, we will address the study of meson spectroscopy in an 
enlarged Hilbert space considering many-quark components. This seems nowadays unavoidable to understand
the experimental data and it builds a bridge towards the description
in terms of hadronic degrees of freedom. Secondly, the same scheme used
to describe the hadron spectroscopy should be valid for describing
the low-energy meson-meson scattering. In other words, contributions to the hadron spectroscopy
arising from many-quark components allowed in the Gell-Mann scheme, should also be recovered
by means of hadron-hadron scattering using a full set of states. Finally, we will review our recent efforts
to describe the structure of heavy baryons containing charm or bottom quarks.

\section{Meson spectroscopy beyond the naive quark-model}
\label{meson}

Many-quark systems present a richer, and therefore more complicated, color structure than standard hadrons
built with a quark-antiquark pair or three quarks. While the color wave function for ordinary mesons and 
baryons leads to a single vector, in the case of four--quark states there are different vectors driving to 
a singlet color state out of either colorless or colored quark-antiquark two-body components. Thus, when 
dealing with four--quark states a basic important question is whether we are in front of a single colorless 
meson--meson molecule or a compact state, defined as a system with two-body colored components. Whereas the 
first structure would be natural in the naive quark model,
the second one would open a new area in hadron spectroscopy.

Let us start by defining a {\it physical channel} as a four-quark state made of two quark-antiquark color singlets.
In Ref.~\cite{Vij09b} the formalism to evaluate the probability of the different physical channels for an
arbitrary four--quark wave function was derived. For this purpose any hidden--color vector of a
four--quark color basis, i.e., vectors with non--singlet internal color couplings,
was expanded in terms of singlet--singlet color vectors. Such a procedure gives rise to a 
wave function expanded in terms of nonorthogonal vectors belonging to different orthogonal basis, and therefore the 
determination of the probability of physical channels becomes cumbersome. Such nonorthogonal 
expansions are common in other fields of science, like chemical physics~\cite{Low50}, 
however they have not been properly discussed until now in the context of the multiquark
phenomenology, i.e., {\it color chemistry}. We have derived 
the two hermitian operators that are well--defined projectors on the  
physical singlet--singlet color states $\ka$ and $\kap$.
Using these operators the probabilities for finding
singlet--singlet color components for an arbitrary four-quark state $|\Psi\rangle$ are given by,
\begin{eqnarray}
P^{\mid\Psi\rangle}({[11]})&=&\frac{1}{2(1-|\,_c\left\langle 11 \mid 1'1' \right\rangle_c|^2)}
\left[ \left\langle\Psi\mid P\hat Q \mid\Psi\right\rangle +
\left\langle\Psi\mid \hat Q P \mid\Psi\right\rangle\right] \nonumber \\
P^{\mid\Psi\rangle}({[1'1']})&=&\frac{1}{2(1-|\,_c\left\langle 11 \mid 1'1' \right\rangle_c|^2)}
\left[ \left\langle\Psi\mid \hat P Q \mid\Psi\right\rangle +
\left\langle\Psi\mid Q \hat P \mid\Psi\right\rangle\right] \, ,
\end{eqnarray}
where $P(\hat P)$ and $Q(\hat Q)$ are the projectors over the different color vectors, 
$\ka(\kap)$ and $\kb(\kbp)$ respectively. 

The stability of a four--quark state relies on $\Delta_E=E_{4q}-E(M_1,M_2)$,
the energy difference between its mass and that of the lowest two-meson threshold. 
It is important to emphasize the relevance of comparing four--quark energies with respect
to thresholds obtained using the same quark-quark interaction as in the four--quark case. 
Only in doing so one may guarantee that no spurious bound states are obtained.
$\Delta_E<0$ indicates that all fall-apart decay channels
are forbidden, and therefore one has a proper bound state. $\Delta_E\ge 0$
will indicate that the four--quark solution corresponds to
an unbound threshold (two free mesons). As we will discuss below, between these
two limits one may find molecular states.

\begin{table}[b]
\begin{center}
\begin{tabular}{|c|ccccc|}
\hline
 $(S,I,L=0)$                       & (0,1)          &  (1,1)          & (1,0)         & (1,0)          & (0,0) \\
Flavor                          &$cc\bar n\bar n$&$cc\bar n\bar n$&$cc\bar n\bar n$&$bb\bar n\bar n$&$bb\bar n\bar n$\\
\hline
Energy (MeV)                          & 3877           &  3952           & 3861          & 10395          & 10948 \\
Threshold                       & $DD\mid_S$     &  $DD^*\mid_S$   & $DD^*\mid_S$   & $BB^*\mid_S$   &  $B_1B\mid_P$\\
$\Delta_E$ (MeV)                      & +5             &  +15            & $-76$         & $-$217         &  $-153$ \\
\hline
$P[| \bar 3 3\rangle_c^{12}]$   & 0.333          &  0.333          & 0.881         & 0.974          &  0.981 \\
$P[| 6 \bar 6\rangle_c^{12}]$   & 0.667          &  0.667          & 0.119         & 0.026          &  0.019 \\
\hline
$P[\ka]$                        & 0.556          &  0.556          & 0.374         & 0.342          &  0.340 \\
$P[\kb]$                        & 0.444          &  0.444          & 0.626         & 0.658          &  0.660 \\
\hline
$P_{MM}$                        & 1.000          &  $-$            & $-$           & $-$            &  0.254 \\
$P_{MM^*}$                      & $-$            &  1.000          & 0.505         & 0.531          &  $-$ \\
$P_{M^*M^*}$                    & 0.000          &  0.000          & 0.495         & 0.469          &  0.746 \\
\hline
\end{tabular}
\caption{Four--quark state properties for selected quantum numbers. The notation
$M_1M_2\mid_{\ell}$ stands for mesons $M_1$ and $M_2$ with relative orbital
angular momentum $\ell$. $P[| \bar 3 3\rangle_c^{12}(| 6\bar 6\rangle_c^{12})]$ and
$P[\ka(\kb)]$ stand for the probability of the color components in different basis~\protect{\cite{Sym09}}.}
\label{t1}
\end{center}
\end{table}
The four--body $QQ\bar n\bar n$ ($Q$ stands for a heavy $c$ or $b$ quark and $n$ for a light $u$, $d$ or $s$ quark) 
Schr\"odinger equation has been solved using two independent methods, the hyperspherical harmonic formalism~\cite{Vij09c}
and a variational approach based on generalized gaussians~\cite{Sym09}.
All possible quantum numbers for both double charm and double bottom four-quark states were evaluated. 
Among all possible combinations, in the charm sector only one bound state with
quantum numbers $(I) J^P = (0) 1^+$ has been found. We show in Table~\ref{t1} a summary of the results obtained for several 
bound and unbound four--quark states in the bottom and charm sectors. One can see how, independently of the binding energy, all of them contain 
a sizable octet-octet color component. Let us first of all concentrate on the two unbound states. Using the formalism of Ref.~\cite{Vij09b} one can
evaluate the probability of the different physical channels: $P_{MM}$, two pseudoscalar mesons, $P_{M^*M^*}$, two vector mesons, and $P_{MM^*}$, 
a pseudoscalar and a vector meson. One can see how four--quark unbound states are represented by two isolated mesons.
Let us now turn to the bound states.
In contrast to the results obtained for unbound states, the probabilities in several of the allowed physical channels are
relevant. Thus, it becomes clear how the bound state must be generated by an interaction
that is not present in the asymptotic channel, sequestering probability
from a single singlet--singlet color vector due to the interaction between
color octets. Such systems are clear examples
of compact four--quark states, in other words, they cannot be expressed in terms of a single physical
channel. 
\begin{figure}[t]
\begin{center}
\vspace*{0.2cm}
\epsfig{file=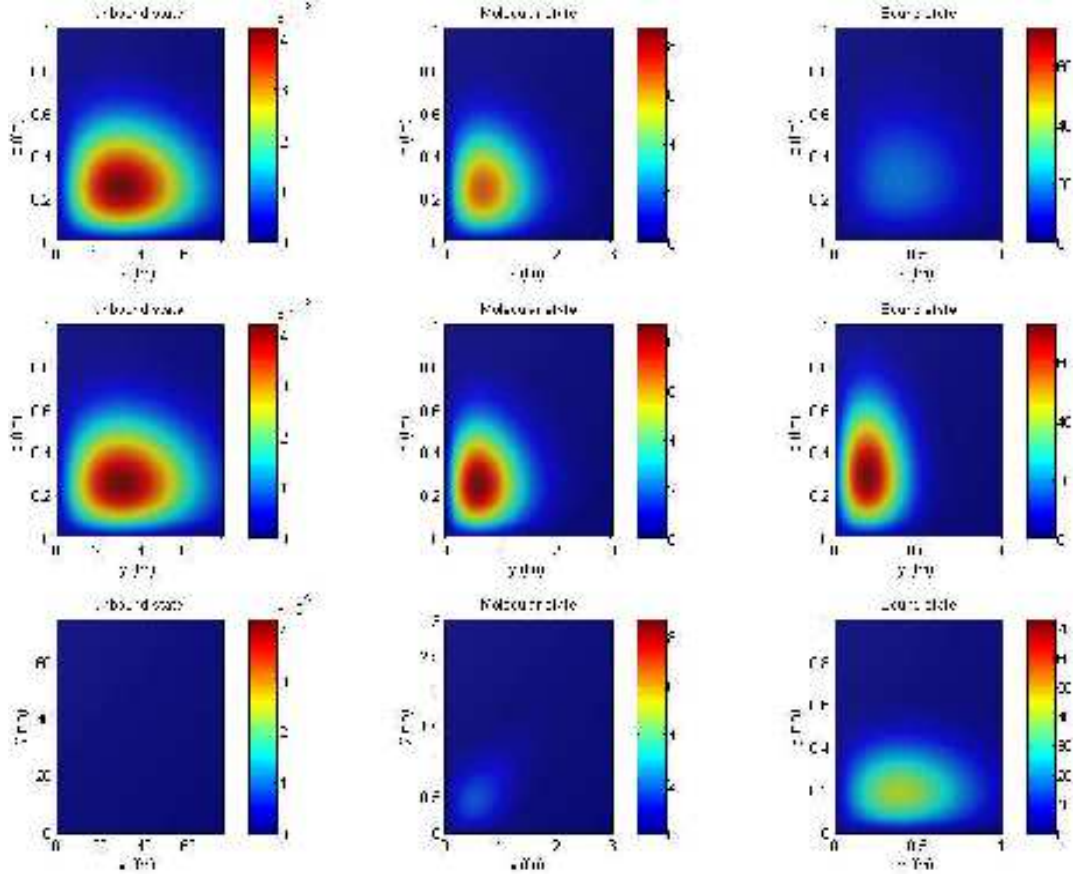, width=14cm}
\caption{Position probability distributions corresponding to a unbound: ($S=1$, $I=1$) $cc\bar n\bar n$,
a molecular: ($S=0$, $I=1$) $cc\bar n\bar n$, and a bound state: ($S=1$, $I=0$) $bb\bar n\bar n$.}
\label{colorin}
\end{center}
\end{figure}

To illustrate in more detail the differences observed in the calculated four--quark wave functions we depict in
Fig.~\ref{colorin} the position probability distributions defined as
\begin{equation}
\label{coloreq}
R(r_{\alpha},r_{\beta})=r_{\alpha}r_{\beta}\sum_i\int_V |R_i(\vec r_{\alpha},\vec r_{\beta},\vec r_{\gamma})|^2 d\vec r_{\gamma}\,d\Omega_{r_{\alpha}}\,d\Omega_{r_{\beta}}\,,
\end{equation}
where $R_i(\vec r_{\alpha},\vec r_{\beta},\vec r_{\gamma})$ are the four--quark radial wave functions. 
We present results for a unbound, a molecular
and a bound state, showing the position distribution for the different planes
$(r_{\alpha},r_{\beta})=(z,x),(z,y)$, and $(x,y)$. Clear differences among them
can be observed. The position distribution for the unbound case spreads in the
$x$ and $y$ Jacobi coordinates up to 60 fm, while the bound and molecular systems are
restricted to the region below 3 fm (molecular) and 1 fm (bound). In the $(x,y)$
plane the unbound state is so widely spread that the values for the position
distribution are three orders of magnitude lower than in the $(z,y)$ and $(z,x)$ cases,
and therefore they will not appear in the picture unless artificially magnified.
In the case of the molecular state a long range tail propagating in the $x=y$ region
can be observed contrary to the constrained values obtained for bound systems.

In a recent investigation, the $Q\overline{Q}n\bar n$  Schr\"odinger equation
has been solved accurately using the hyperspherical harmonic formalism~\cite{Vij07}. 
All possible quantum numbers were discussed by means of different constituent quark models widely used in the literature.
No compact bound states were found for any set of quantum numbers nor any constituent quark model. 
Thus, independently of the quark--quark interaction and the quantum numbers 
considered, the system evolves to a well separated two-meson state. In any manner one can claim for the existence
of a compact bound state for the $c\bar c n \bar n$ system. 

Unfortunately, close to a threshold, methods based on a series expansion fail to converge
since arbitrary large number of terms are required to determine the wave function.
For this reason, we have solved the two-body Lippmann-Schwinger equation 
for negative energies using the Fredholm determinant method,
looking for attractive channels that may contain a meson-meson molecule,
that permitted us to obtain robust predictions even
for zero-energy bound states~\cite{PRL09}. 
To do so we have started from a physical system made of two mesons, $M_1$ and $M_2$, with quantum numbers $(I)J^{PC}$ in a relative 
$S-$wave. Then, in general, there is a coupling to the $M_1 M_2$ $D-$wave and to any other two--meson system that can couple to 
the same quantum numbers $(I)J^{PC}$. Thus, the Lippmann-Schwinger equation for the $M_1M_2$ scattering becomes
\begin{equation}
t_{\alpha\beta;JI}^{\ell_\alpha s_\alpha, \ell_\beta s_\beta}(p_\alpha,p_\beta;E) =  
V_{\alpha\beta;JI}^{\ell_\alpha s_\alpha, \ell_\beta s_\beta}(p_\alpha,p_\beta)+ \sum_{\scriptstyle \gamma}\sum_{\ell_\gamma=0,2} 
\int_0^\infty p_\gamma^2 dp_\gamma V_{\alpha\gamma;JI}^{\ell_\alpha s_\alpha, \ell_\gamma s_\gamma}
(p_\alpha,p_\gamma) G_\gamma(E;p_\gamma)
t_{\gamma\beta;JI}^{\ell_\gamma s_\gamma, \ell_\beta s_\beta}
(p_\gamma,p_\beta;E)\, ,
\label{eq0}
\end{equation}
where $\alpha \equiv \beta \equiv M_1M_2$, $\gamma$ stands for any intermediate two-body state that can couple
to the $(I)J^{PC}$ quantum numbers, $t$ is the two-body scattering amplitude, $J$, $I$, and $E$ are the angular momentum, isospin 
and energy of the two-body system,
$\ell_{\alpha} s_{\alpha}$, $\ell_{\gamma} s_{\gamma}$, and $\ell_{\beta} s_{\beta }$ are the initial, intermediate, and final 
orbital angular momentum and spin, respectively,  and $p_\gamma$ is the relative momentum of the two-body system $\gamma$. 
To solve the scattering problem we have derived a meson-meson potential in the Born--Oppenheimer
approach from the basic $\bar{q} q$ interaction used to study the four--quark system with the 
hyperspherical harmonic formalism or the generalized gaussian variational approach.

\begin{figure}[t]
\begin{center}
\includegraphics[scale=0.4]{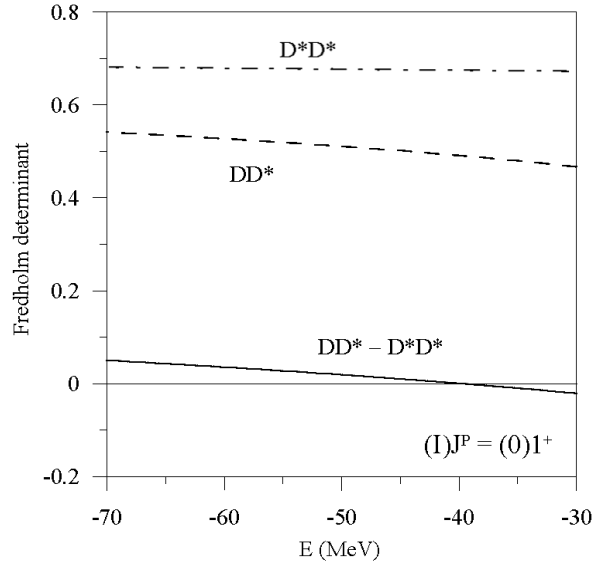}
\end{center}
\caption{Fredholm determinant for the $(I)J^{P}=(0)1^{+}$ $cc\bar n\bar n$
system. The dashed line corresponds to the calculation considering only
$DD^*$ singlet-singlet color states. The dashed--doted line stands for the results considering only the $D^*D^*$ system. The solid line represents the results
including the coupling $DD^*\leftrightarrow D^*D^*$.}
\label{fredholm}
\end{figure}

We have analyzed all positive--parity channels made by $S-$wave interacting mesons 
up to $J^P = 2^+$ in the $QQ\bar n\bar n$ sector. Only the channel $(I) J^P = (0) 1^+$
is attractive enough to be bound~\cite{Carames:2011zz}. This is the same state that was found to be bound in the four--quark study. 
As can be seen in Fig.~\ref{fredholm} the $DD^*$ and $D^*D^*$ potentials are attractive, however none of them is bound
by itself. A bound state is obtained when one considers the coupling between the $DD^*$ and the $D^*D^*$ systems and solves 
the coupled--channel problem accordingly. This mixing at short
distances will reconstruct the octet-octet color vector component of a particular basis as shown in Table~\ref{t1}. 

When moving to the bottom sector one finds four more candidates for observation. They are, ordered by their 
binding energy: $(I) J^P = (0) 1^+$, $(0)$ $0^+$, $(1)$ $3^-$ and $(0)$ $1^-$. 
The interest in these states is increasing as for the first time there are chances to observe such large mass exotic states 
in quite a near future. LHC may discover tetraquark states via gluon--gluon fusion due to 
both large number of events and their unique signature in the 
detectors~\cite{Yuqi:2011gm}. RHIC might identify hadronic molecular states by means of relativistic heavy ion collisions 
by employing the coalescence model for hadron production~\cite{Cho:2010db}. If any of these suggestions gets real, 
the new data related to these double charm or bottom exotic systems will have a huge impact on heavy quark spectroscopy. 

In the $Q\overline{Q}n\bar n$ case, $D\bar D$ and charmonium-light two-mesons systems get coupled.
A few channels are slightly attractive (Fredholm determinant smaller than one),
$D\overline{D}$ with $(I)J^{PC}=(0)0^{++}$, $D\overline{D}^*$ with $(0)1^{++}$ and
$D^*\overline{D}^*$ with $(0)0^{++}$, $(0)2^{++}$, and $(1)2^{++}$.
The remaining ones are either repulsive or have zero probability to contain a bound state or a resonance.
The only bound state appears in the $(I)J^{PC}=(0)1^{++}$ channel as a consequence of the 
coupling between $D\bar D^*$ and $J/\Psi \omega$ two-meson channels. It has
a binding energy in the range $0-1$ MeV, and therefore in agreement with the experimental measurements of the $X(3872)$.
The corresponding isovector channel becomes repulsive due to the coupling to the lightest channel that 
includes a pion. Therefore, the existence of charged partners for the $X(3872)$ can be discarded as well
as a partner in the bottom sector~\cite{Ter12}.

Although the last analysis has been performed by means of a particular
quark interacting potential, the conclusions derived are general. 
They mainly rely on using the same hamiltonian to describe tensors of
different order, two-- and four--quark components in the present case. When dealing with a
complete basis, any four--quark deeply bound state has to be compact. Only
slightly bound systems could be considered as molecular. Unbound states correspond
to a two--meson system. A similar situation would
be found in the two baryon system, the deuteron could be considered as a 
molecular--like state with a small percentage of its wave function on the $\Delta \Delta$ channel,
whereas the $H-$dibaryon would be a compact six--quark state.
When working with central forces, the only way of getting a bound system is to have
a strong interaction between the constituents that are far apart in the asymptotic limit
(quarks or antiquarks in the present case). In this case the short--range
interaction will capture part of the probability of a two--meson threshold to form a bound
state. This can be reinterpreted as an infinite sum over physical states.
This is why the analysis performed here is so important before any conclusion can be made concerning 
the existence of compact four--quark states beyond simple molecular structures.

Lets conclude this section by emphasizing that at short distances identical quarks can 
recouple to different vectors of the Hilbert space. If any of them is not considered, a 
fundamental ingredient of the calculation would be neglected. Such effect would never happen 
when dealing with hadronic degrees of freedom. Therefore, the need to incorporate the complete 
Hilbert space is evident in both approaches. If the model 
parameters were fitted to some observables they would necessarily include the effect of a 
restricted Hilbert space and one should never use them for different quantum numbers without 
arriving to wrong conclusions, loosing thus any predictive power. We can therefore conclude that formalisms 
based on four--quark and meson--meson configurations are fully compatible whenever 
all the relevant basis vectors are taken into account. 

\section{Heavy baryon spectroscopy}
\label{baryon}

Heavy baryons containing a single heavy quark are particularly
interesting. The light degrees of freedom (quarks and gluons)
circle around the nearly static heavy quark. Such a system behaves as
the QCD analogue of the familiar hydrogen bounded by the 
electromagnetic interaction. 
When the heavy quark mass $m_Q \to \infty$, the angular
momentum of the light degrees of freedom is a good 
quantum number. Thus, heavy quark baryons belong to 
either SU(3) antisymmetric $\mathbf{\bar{3}_F}$ or symmetric 
$\mathbf{6_F}$ representations. The spin of the light
diquark is 0 for $\mathbf{\bar{3}_F}$, while it is 1
for $\mathbf{6_F}$. Thus, the spin of the ground
state baryons is $1/2$ for $\mathbf{\bar{3}_F}$, representing
the $\Lambda_b$ and $\Xi_b$ baryons, while it can be
both $1/2$ or $3/2$ for $\mathbf{6_F}$, allocating
$\Sigma_b$, $\Sigma^*_b$, $\Xi'_b$, $\Xi^*_b$, $\Omega_b$
and $\Omega^*_b$, where the star indicates spin $3/2$ states.
Therefore heavy hadrons form doublets.
For example, $\Sigma_b$ and $\Sigma_b^*$ will be degenerate
in the heavy quark limit, being their mass splitting caused
by the chromomagnetic interaction at the order $1/m_Q$.
These effects can be, for example, taken into account systematically in the
framework of heavy quark effective field theory.
The mass difference between states belonging to the 
$\mathbf{\bar{3}_F}$ and $\mathbf{6_F}$ representations
do also contain the dynamics of the light diquark 
subsystem, hard to accommodate in any heavy quark mass
expansion. Therefore, exact solutions of the three-body
problem for heavy hadrons are theoretically
desirable because they will serve to test the reliability   
of approximate techniques: heavy quark mass expansions, 
variational calculations, or quark-diquark approximations.

The combined study of $Qnn$ and $QQn$ systems will also provide 
some hints to learn about the basic dynamics governing the interaction
between light quarks. The interaction between
pairs of quarks containing a heavy quark $Q$ is driven by the 
perturbative one-gluon exchange. For the $Qnn$
system the mass difference between members of the  
$\mathbf{6_F}$ SU(3) representation comes determined only
by the perturbative one-gluon exchange, while between members
of the $\mathbf{6_F}$ and $\mathbf{\bar{3}_F}$ representations
it presents contributions from the one-gluon and pseudoscalar exchanges. If the latter mass difference
is attributed only to the one-gluon exchange (this would be the case
of models based only on the perturbative one-gluon exchange), it will be strengthened
as compared to models considering pseudoscalar potentials at the 
level of quarks, where a weaker one-gluon exchange will play the role.
When moving to the $QQn$ systems only one-gluon exchange interactions
between the quarks will survive, with the strength determined in the
$Qnn$ sector, where we have experimental data. This will give rise
to larger masses for the ground states, due to the more attractive 
one-gluon exchange potential in the $Qnn$ sector, what requires
larger constituent quark masses to reproduce the experimental data.

In Ref.~\cite{Val08} the three-body Schr\"odinger equation was solved by the 
Faddeev method in momentum space with the constituent
quark model of Ref.~\cite{Vij05}.
The results are shown in Table \ref{t12} compared to 
experiment and other theoretical approaches. 
All known experimental data are nicely described. 
Such an agreement and the exact method used to solve the three-body problem make our
predictions also valuable as a guideline to experimentalists.
\begin{table}
\begin{center}
\begin{tabular}{|cc||ccccc|ccccc|}
\hline
\multicolumn{2}{|c||}{} &
\multicolumn{5}{|c|}{Charm} &
\multicolumn{5}{|c|}{Bottom} \\
State & $J^P$ & \protect\cite{Val08} & Exp.  & 
\protect\cite{Sil96} &
\protect\cite{Rob07} &
\protect\cite{Ebe08} & 
\protect\cite{Val08} & 
Exp.  & 
\protect\cite{Sil96} &
\protect\cite{Rob07} &
\protect\cite{Ebe08} \\
\hline
$\Lambda_i$  &  $1/2^+$ & 2285 & 2286 & 2285 & 2268 & 2297  
                        & 5624 & 5619 & 5638 & 5612 & 5622 \\
             &  $1/2^+$ & 2785 & 2765 & 2865 & 2791 & 2772 
                        & 6106 &      & 6188 & 6107 & 6086 \\
             &  $1/2^-$ & 2627 & 2595 & 2635 & 2625 & 2598  
                        & 5947 & 5912  & 5978 & 5939 & 5930 \\
             &  $3/2^+$ & 3061 &      & 2930 & 2887 & 2874  
                        & 6388 &      & 6248 & 6181 & 6189 \\ \hline
$\Sigma_i$   &  $1/2^+$ & 2435 & 2454 & 2455 & 2455 & 2439  
                        & 5807 & 5811 & 5845 & 5833 & 5805 \\
             &  $1/2^+$ & 2904 &      & 3025 & 2958 & 2864  
                        & 6247 &      & 6370 & 6294 & 6202 \\    
             &  $1/2^-$ & 2772 & 2765 & 2805 & 2748 & 2795  
                        & 6103 &      & 6155 & 6099 & 6108 \\
             &  $3/2^+$ & 2502 & 2518 & 2535 & 2519 & 2518  
                        & 5829 & 5833 & 5875 & 5858 & 5834 \\ \hline          
$\Xi_i$      &  $1/2^+$ & 2471 & 2471 & 2467 & 2492 & 2481  
                        & 5801 & 5793 & 5806 & 5844 & 5812 \\
             & $1/2'^+$ & 2574 & 2578 & 2567 & 2592 & 2578  
                        & 5939 &      & 5941 & 5958 & 5937 \\
             &  $1/2^-$ & 2799 & 2792 & 2792 & 2763 & 2801  
                        & 6109 &      & 6116 & 6108 & 6119 \\
             &  $3/2^+$ & 2642 & 2646 & 2647 & 2650 & 2654  
                        & 5961 & 5945 & 5971 & 5982 & 5963 \\ \hline                        
$\Omega_i$   &  $1/2^+$ & 2699 & 2698 & 2675 & 2718 & 2698  
                        & 6056 & 6054 & 6034 & 6081 & 6065 \\
             &  $1/2^+$ & 3159 &      & 3195 & 3152 & 3065 
                        & 6479 &      & 6504 & 6472 & 6440 \\
             &  $1/2^-$ & 3035 &      & 3005 & 2977 & 3020  
                        & 6340 &      & 6319 & 6301 & 6352 \\
             &  $3/2^+$ & 2767 & 2768 & 2750 & 2776 & 2768 
                        & 6079 &      & 6069 & 6102 & 6088 \\ \hline
\end{tabular}
\caption{Masses, in MeV, of charmed and bottom baryons.}
\label{t12}
\end{center}
\end{table}

It is worth noticing that the relativistic quark-diquark approximation~\cite{Ebe08}
and the harmonic oscillator variational method~\cite{Rob07} predict a lower
$3/2^+$ excited state for the $\Lambda_b$ baryon.
Such result can be easily understood by the influence  of the pseudoscalar 
interaction between the light quarks on the 
$\Lambda_b(1/2^+)$ ground state. If this attraction were not present for the $\Lambda_b(1/2^+)$, the 
$\Lambda_b(3/2^+)$ would be lower in mass as reported
in Refs.~\cite{Rob07,Ebe08} (a similar effect will be 
observed in the charmed baryon spectra).
Thus, the measurement and identification of the 
$\Lambda_b(3/2^+)$ is a relevant feature that will help to
clarify the nature of the interaction between the light
quarks in heavy baryon spectroscopy.

Finally, we can make parameter free predictions for double charmed and bottom baryons. 
Our results are shown in Table~\ref{t20}. For double charmed baryons,
the ground state is found to be at 3579 MeV, far below the result of
Ref.~\cite{Rob07} and in perfect agreement with lattice nonrelativistic
QCD~\cite{Mat02}, but still a little bit higher than the 
non-confirmed SELEX result, 3519 MeV~\cite{Och05}. It is therefore
a challenge for experimentalists to confirm or to find the ground state
of double charmed and bottom baryons. 
\begin{table}[t]
\begin{center}
\begin{tabular}{|cc|ccccc|}
\hline
State  & 
$J^P$ & 
\protect\cite{Val08} & 
\protect\cite{Ron95} & 
\protect\cite{Sil96} & 
\protect\cite{Mat02} & 
\protect\cite{Rob07} \\  
\hline
$\Xi_{bb}$      & $1/2^+$     &10189&10340&10194&     & 10340 \\ \hline
    $\Delta E$  & $3/2^+$     &  29 &  30 &  41 &  20 &  27  \\     
                & $1/2^-$     & 217 &     & 262 &     & 153  \\    \hline
$\Omega_{bb}$   & $1/2^+$     &10293&10370&10267&     & 10454  \\\hline
 $\Delta E$      & $3/2^+$     &  28 &  30 &  38 &  19 &  32  \\ 
                & $1/2^-$     & 226 &     & 265 &     & 162  \\  \hline
$\Xi_{cc}$      & $1/2^+$     & 3579& 3660& 3607& 3588& 3676  \\  \hline     
 $\Delta E$     & $3/2^+$     &  77 &  80 &  93 &  70 &  77  \\  
                & $1/2^-$     & 301 &     & 314 &     & 234  \\  \hline
$\Omega_{cc}$   & $1/2^+$     & 3697& 3740& 3710& 3698& 3815  \\  \hline          
 $\Delta E$     & $3/2^+$     &  72 &  40 &  83 &  63 &  61  \\  
                & $1/2^-$     & 312 &     & 317 &     & 231  \\ \hline
\end{tabular}
\caption{Ground state and excitation energies, $\Delta E$, of double charmed and bottom
baryons. Masses are in MeV.}
\label{t20}
\end{center}
\end{table}

\section{Summary}
\label{summary}

To summarize this talk, we have reviewed some recent results of hadron spectroscopy
and hadron-hadron interaction from the point of view of the constituent quark model. 
The constituent quark model is a powerful tool at our disposal to face the 
new issues appearing in an almost daily basis in the heavy quark sector. Unfortunately, the 
downsides of such robust tools lie in the risk of misusing them.
Therefore, we have tried to impress upon the reader the range of validity and proper formalisms
of the constituent quark model. Similarly, since there are many different models in the literature, 
we have tried to present conclusions that are independent of the particularities of the model chosen.
We have seen how the enlargement of the Hilbert space when increasing energy, that was seen to be necessary
to describe the nucleon-nucleon phenomenology above the pion production
threshold, seems to be necessary to understand current problems of hadron
spectroscopy. We have also tried to emphasize that spectroscopy and interaction
can, and must, be understood within the same scheme when dealing with quark
models. Any other alternative becomes irrelevant from the point of view
of learning about properties of QCD.

It is expected that in the coming years better-quality data on all sectors
discussed in this talk  will become available and our results can be used
to analyze these upcoming data in a model--independent way. 

\section*{Acknowledgments}
This work has been partially funded by the Spanish Ministerio de
Educaci\'on y Ciencia and EU FEDER under Contract No. FPA2010-21750,
and by the Spanish Consolider-Ingenio 2010 Program CPAN (CSD2007-00042).


\begin{thebibliography}{99}

\bibitem{Bjo85} J.~D.~Bjorken, The November Revolution: A Theorist Reminisces, 
in: A Collection of Summary Talks in High Energy Physics (ed. J.D.~Bjorken),
p. 229 (World Scientific, New York, 2003).

\bibitem{Kel01} W.~T.~Kelvin, 
	      The London, Edinburgh and Dublin Philosophical Magazine and Journal of Science, Series 6, volume 2, page 1 (1901).

\bibitem{Gel64} M.~Gell-Mann,
				Phys. Lett. {\bf 8} (1964) 214-215. 

\bibitem{Val05} A.~Valcarce, H.~Garcilazo, F.~Fern\'andez, and P.~Gonz\'alez,
				Rep. Prog. Phys. {\bf 68} (2005) 965-1042.			

\bibitem{Vij09b} J.~Vijande and A.~Valcarce
        Phys. Rev. C {\bf 80} (2009) 035204(1-10).

\bibitem{Low50} P.~-O.~L\"owdin, 
				J. Chem. Phys. {\bf 18} (1950) 365-376.

\bibitem{Vij09c} J.~Vijande, A.~Valcarce, and N.~Barnea,
                Phys. Rev. D {\bf 79} (2009) 074010(1-16).

\bibitem{Sym09} J.~Vijande and A.~Valcarce,
								Symmetry {\bf 1} (2009) 155-179.
 
\bibitem{Vij07} J.~Vijande, E.~Weissman, N.~Barnea, and A.~Valcarce,
                Phys. Rev. D {\bf 76} (2007) 094022(1-17).

\bibitem{PRL09} T.~Fern\'andez-Caram\'es, A.~Valcarce, and J.~Vijande,
                Phys. Rev. Lett. {\bf 103} (2009) 222001(1-4).
              
\bibitem{Carames:2011zz}  T.~F.~Caram\'es, A.~Valcarce and J.~Vijande,
		Phys. Lett. B {\bf 699} (2011) 291-295.
  
\bibitem{Yuqi:2011gm}  C.~Yu-qi and W.~Su-zhi,
		Phys. Lett. B {\bf 705} (2011) 93-97.

\bibitem{Cho:2010db} S.~Cho {\it et al.} [ExHIC Collaboration],
		Phys. Rev. Lett. {\bf 106} (2011) 212001(1-4). 
 
\bibitem{Ter12} T.~F.~Caram\'es, A.~Valcarce and J.~Vijande,
		Phys. Lett. B {\bf 709} (2012) 358-361.

\bibitem{Val08} A.~Valcarce, H.~Garcilazo, and J.~Vijande.
		Eur. Phys. J. A{ \bf 37} (2008) 217-225.
		
\bibitem{Vij05} J.~Vijande, F.~Fern\'andez, and A.~Valcarce,
                J. Phys. G {\bf 31} (2005) 481-512.

\bibitem{Sil96} B.~Silvestre-Brac,
		Few-Body Systems {\bf 20} (1996) 1-25.

\bibitem{Rob07} W.~Roberts and M.~Pervin,
		Int. J. Mod. Phys. A{\bf 23} (2008) 2817-2860.

\bibitem{Ebe08} D.~Ebert, R.~N.~Faustov, and V.~O.~Galkin, 
		Phys. Lett. B {\bf 659} (2008) 612-620.
 
\bibitem{Mat02} N.~Mathur, R.~Lewis, and R.~M.~Woloshyn, 
		Phys. Rev. D{\bf 66} (2002) 014502(1-10).

\bibitem{Och05} A.~Ocherashvili {\it et al.} [SELEX Collaboration],
		Phys. Lett. B {\bf 628} (2005) 18-24.

\bibitem{Ron95} R.~Roncaglia, D.~B.~Lichtenberg, and E.~Predazzi,
		Phys. Rev. D {\bf 52} (1995) 1722-1725.

\end{thebibliography}
\end{document}